\documentclass[11pt,twoside,smallheadings,a4paper]{scrartcl}
\usepackage[margin=1in,nohead,bindingoffset=0cm,footskip=2.5\baselineskip
]{geometry}

\newcommand{\HowPublished}{{}}
\usepackage{url,xspace,floatflt,graphicx}
\usepackage[tight,hang]{subfigure}
\usepackage{pslatex,clrscode
  ,kb}
\usepackage[deco]{myheader}
\usepackage{
  mytheorem}

\newcommand{\natplus}{\nat_+}

\newcommand{\Lan}{{\cal L}}

\newcommand{\omitlay}{}

\newcommand{\EmptyProg}{\epsilon}

\newcommand{\dotty}[1][,]{#1\ldots #1}

\newcommand{\agents}{\mathbb{P}}

\newcommand{\snd}[2][]{\itext{snd}_{#1}(#2)}

\newcommand{\rcv}[2][]{\itext{rcv}_{#1}(#2)}

\newcommand{\lacause}{\xrightarrow{\scriptscriptstyle\text{\textsc{l}}}}
\newcommand{\chan}[2]{\sftext{chan}_{#1\to #2}}

\newcommand{\valt}[2][1\to 2]{%
  \sftext{val\_transmit}_{\substack{#1\\#2}}}

\newcommand{\rel}{\sctext{Rel}\xspace}

\newcommand{\relfifo}{\sctext{RelFi}\xspace}

\renewcommand{\Var}{\itext{Var}}

\newcommand{\from}{\leftarrow}

\newcommand{\thesnd}{e_{\textsc{s}}}
\newcommand{\thercv}{e_{\textsc{r}}}
\newcommand{\chopname}{silent cut\xspace}
\newcommand{\chop}{\mathbin{\wr}}

\title{Causing Communication Closure:
  Safe Program Composition with Reliable Non-FIFO Channels%
  \thanks{A preliminary version appeared as~\cite{EM2005a}. Work was
    partially supported by ARC Discovery Grant RM02036.}}

\author{Kai Engelhardt%
  \thanks{
    \texttt{kaie@cse.unsw.edu.au}, School of Computer Science
    and Engineering, The University of New South Wales, and NICTA,
    Sydney, NSW 2052, Australia. National ICT Australia is funded
    through the Australian Government's \textit{Backing Australia's
      Ability} initiative, in part through the Australian Research
    Council.}
  \and Yoram Moses%
  \thanks{\texttt{moses@ee.technion.ac.il}, Department of Electrical
    Engineering, Technion, Haifa, 32000 Israel. Work on this paper
    happened during a sabbatical visit to the School of Computer
    Science and Engineering, The University of New South Wales,
    Sydney, NSW 2052, Australia.}}

\newcommand{\Expr}{\itext{Expr}}
\newcommand{\expr}{\sftext{e}}
\newcommand{\guard}[1]{[#1]}
\newcommand{\refines}{\leq}
\newcommand{\ud}{^{\text{u}}}
\newcommand{\comp}[1]{\overline{#1}}
\newcommand{\tcc}{tail communication closed\xspace}

\newcommand{\TCC}{TCC\xspace}
\newcommand{\BSL}{BSL\xspace}
\newcommand{\BSLs}{BSLs\xspace}

\newcommand{\Id}{\ntext{id}}
\newcommand{\Fd}{\const{fst}}
\newcommand{\Ld}{\const{lst}}
\newcommand{\ack}{\const{ack}}
\renewcommand{\snd}[2][i\to j]{\proc{snd}^{#1}_{#2}}
\renewcommand{\rcv}[2][j\from i]{\proc{rcv}^{#1}_{#2}}
\renewcommand{\valt}[2][i\to j]{\proc{mt}^{#1}_{#2}}
\newcommand{\cardP}{
  n}
\newcommand{\size}[1]{\left\|#1\right\|}
\newlength{\grafflecm}
\newcommand{\goodgap}{%
  \hspace{1.5\grafflecm}}

\newcommand{\valtji}{\mbox{$\valt[j\to i]{}$}}
\newcommand{\sig}{\proc{Sig}}

\newcommand{\withappendix}[1]{}
\begin{document}
\allowdisplaybreaks

\maketitle
\begin{abstract}
  A semantic framework for analyzing safe composition of distributed
  programs is presented.
  Its applicability is illustrated by a study of program composition
  when communication is reliable but not necessarily FIFO\@. In this
  model, special care must be taken to ensure that messages do not
  accidentally overtake one another in the composed program. We show
  that barriers do not exist in this model. Indeed, no program that
  sends or receives messages can automatically be composed with
  arbitrary programs without jeopardizing their intended behavior.
  Safety of composition becomes context-sensitive and new tools are
  needed for ensuring it. A notion of \emph{sealing} is defined, where
  if a program~$P$ is immediately followed by a program~$Q$ that seals
  $P$ then $P$ will be communication-closed---it will execute as if it
  runs in isolation. The investigation of sealing in this model
  reveals a novel connection between Lamport causality and safe
  composition. A characterization of sealable programs is given, as
  well as efficient algorithms for testing if~$Q$ seals~$P$ and for
  constructing a seal for a significant class of programs. It is shown
  that every sealable program that is open to interference on $O(n^2)$
  channels can be sealed using $O(n)$ messages.
\end{abstract}

\allowdisplaybreaks

\maketitle

\section{Introduction}
\label{sec:intro}

Much of the distributed algorithms literature is devoted to solutions
for individual tasks. Implicitly it may appear that these solutions
can be readily combined to create larger applications. Composing such
solutions is not, however, automatically guaranteed to maintain their
correctness and their intended behavior.
For example, algorithms are typically designed under the assumption
that they begin executing in a well-defined initial global state in
which all channels are empty. In most cases, the algorithms are not
guaranteed to terminate in such a state.
Another inherent feature of distributed systems is that, even though
they are often designed in clearly separated phases, these phases
typically execute concurrently. For instance, Lynch writes
in~\cite[p.~523]{Lynch96}:
\begin{quote}
  ``An MST algorithm can be used to solve the leader-election problem
  [\ldots]. Namely, after establishing an MST, the processes
  participate in the \emph{STtoLeader} protocol to
  select the leader.
  Note that the processes do not need to know when the MST algorithm
  has completed its execution throughout the network; it is enough for
  each process $i$ to wait until it is finished locally, [\ldots].''
\end{quote}
In general, when two phases, such as implementations of an MST
algorithm and of the \emph{STtoLeader} algorithm, are developed
independently and then executed in sequence, one phase may confuse
messages originating from the other with its own messages. Perhaps the
first formal treatment of this issue was via the notion of
\emph{communication-closed layers} introduced by Elrad and Francez
in~\cite{Elrad.Francez82:scp}. Consider a program
$P=P_1\parallel\ldots\parallel P_n$ consisting of $n$ concurrent
processes $P_i = Q_i ; L_i ; Q_i'$, the execution of which is,
intuitively, divided into three phases, $Q_i$, $L_i$, and $Q_i'$.
Elrad and Francez define $L = L_1\parallel\ldots\parallel L_n$ to be a
\emph{communication-closed layer (CCL) in $P$} if under no execution
of $P$ does a command in some $L_i$ communicate with a command in any
$Q_j$ or $Q_j'$~\cite{Elrad.Francez82:scp}. If a program $P$ can be
decomposed into a sequence of CCLs then every execution of $P$ can be
viewed as a concatenation of executions of $P$'s layers in order.
Hence, reasoning about $P$ can be reduced to reasoning about its
layers in isolation. This approach has been investigated further and
applied to a variety of problems by Janssen, Poel, and
Zwiers~\cite{JanssenPoelZwiers1991:CONCUR,Janssen1994:PhD,Janssen95:TACAS,PoelZwiers1992:CAV}.
Stomp and de Roever considered related notions in the context of
synchronous communication~\cite{StompdeRoever1994:facs}. Gerth and
Shrira considered the issue of using distributed programs as
off-the-shelf components to serve as layers in larger distributed
programs~\cite{GerthShrira1986:FSTTCS}. They observe that the above
definition of CCL is made with respect to the whole program $P$ as
context, and hence is unsuitable for off-the-shelf components. They
solve the problem by defining $L$ to be a \emph{General Tail
  Communication Closed (GTCC)} layer if, roughly speaking, for
\emph{all} layers $T_1\parallel\ldots\parallel T_n$ we have that $L$
is a CCL in $L_1;T_1\parallel\ldots\parallel L_n;T_n$. Since this
definition does not refer to the surrounding program context of a
layer, it asserts a certain quality of composability. Sequentially
composing GTCC layers guarantees that each one of them is a CCL.

We develop a framework for defining and reasoning about various
notions central to the design of CCLs in different models of
communication.
The communication model used in most of the literature concerning CCLs
is that of reliable FIFO channels.
In practice, channels often fail to satisfy this assumption. Three
main sources of imperfection are loss, reordering, and duplication of
messages by a channel.
This paper studies the impact of message reordering on the design of
CCLs. Our communication model, which we call \rel, will therefore
assume that channels neither lose nor duplicate messages but message
delivery is not necessarily FIFO.
As we shall see, in \rel, the CCL property depends in an essential way
on Lamport causality~\cite{Lamport1978:cacm}. Indeed, to ensure CCL,
causality is \emph{all} that is needed in \rel, whereas either
duplication or loss already mandate the need for  headers in
messages~\cite{FeketeLynch1990:CONCUR,EM2005c}.


\begin{floatingfigure}[l]{9.3\grafflecm}
  \begin{center}
    \includegraphics[width=9.3\grafflecm]{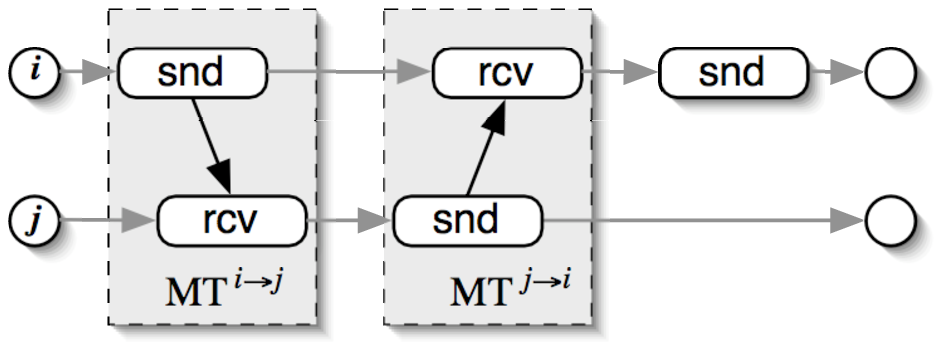}%
    \caption{$\valtji$ seals $\valt{}$.}
    \label{fig:vtseals}
  \end{center}
\end{floatingfigure}

Consider for instance the task of transmitting a message $m$ from
process $i$ to process $j$ where it is stored in variable $x$. The
task is accomplished by $i$ performing $\snd[i\to j]{m}$ to send the
message and $j$ performing $\rcv[j\from i]{x}$ to receive it into
variable $x$. This implementation denoted $\valt{m\to x}$ (for
\emph{Message-Transmit}) works fine in isolation. Composing two
copies\footnote{We omitted the subscript in $\valt{}$. Whenever a
  parameter is irrelevant to the point being made, we tend to omit
  it.} of $\valt{}$, however, does not guarantee the same behavior as
executing the first to completion and then executing the second. Since
communication is not FIFO, the second message sent by $i$ could be the
first one received by $j$.
On the other hand, if $\valt{}$ is followed by $\valt[j\to i]{}$ no
such interference occurs. Moreover, no later program can ever
interfere with the first $\valt{}$ in this pair. Of course the second
program, $\valt[j\to i]{}$, is still susceptible to interference,
e.g., by another $\valt[j\to i]{}$.
In fact, non-trivial programs are never safe from interference in
\rel. As we shall show, for any terminating program $P$ transmitting a
message from $i$ to $j$ there is a program $Q$ potentially interfering
with communication in $P$. One consequence is that no terminating
program that sends messages can be a GTCC layer.

\begin{figure}[thbp]
  \begin{center}
    \subfigure[A program $P$.
    \label{fig:3sealP}]{%
      \includegraphics[width=10.6\grafflecm]{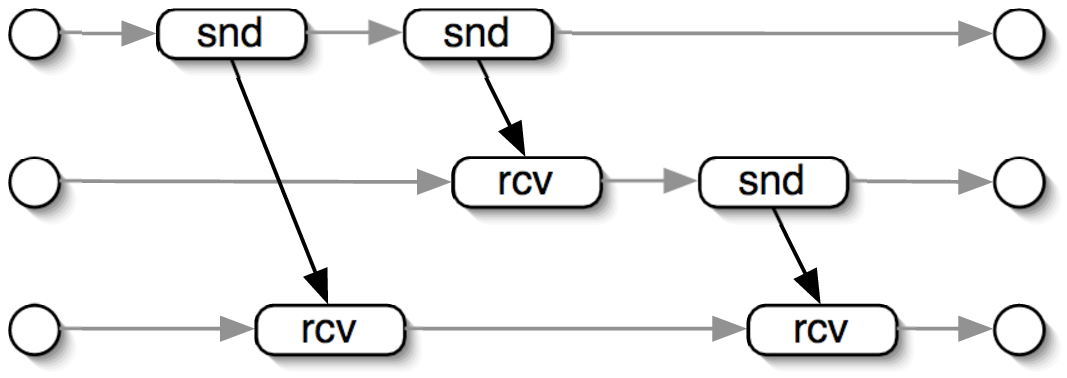}}\goodgap
    \subfigure[A seal for $P$.
    \label{fig:3sealS}]{%
      \includegraphics[width=8.0\grafflecm]{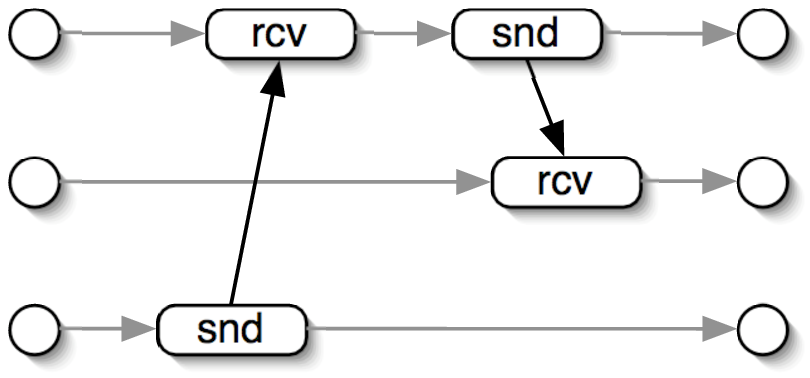}}
  \end{center}
  \caption{An example of sealing.}
  \label{fig:3seal}
\end{figure}

\begin{figure}[thbp]
  \begin{center}
    \subfigure[Unsealable program $X$.
    \label{fig:unseal}]{%
      \includegraphics[width=7.1\grafflecm]{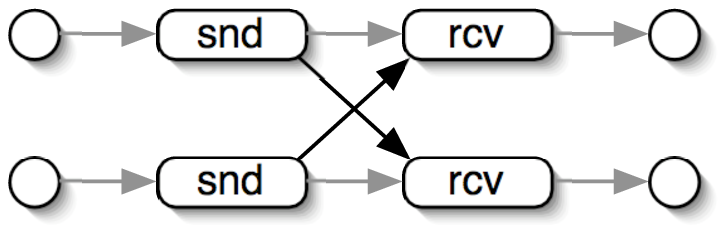}}\\
    \subfigure[A sealable program $P'$.
    \label{fig:unsealP}]{%
      \includegraphics[width=7.1\grafflecm]{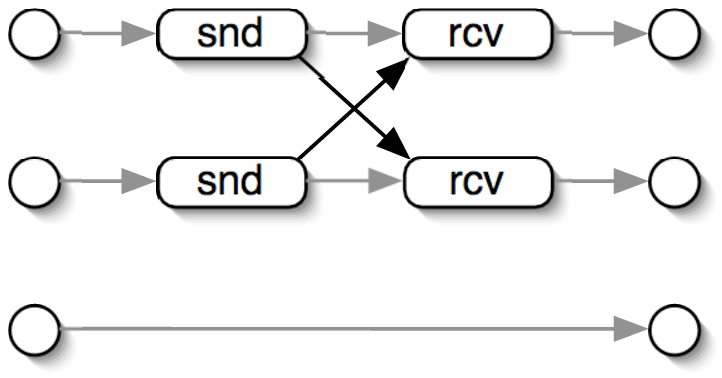}}\goodgap
    \subfigure[A seal for $P'$.
    \label{fig:unsealS}]{%
      \includegraphics[width=10.8\grafflecm]{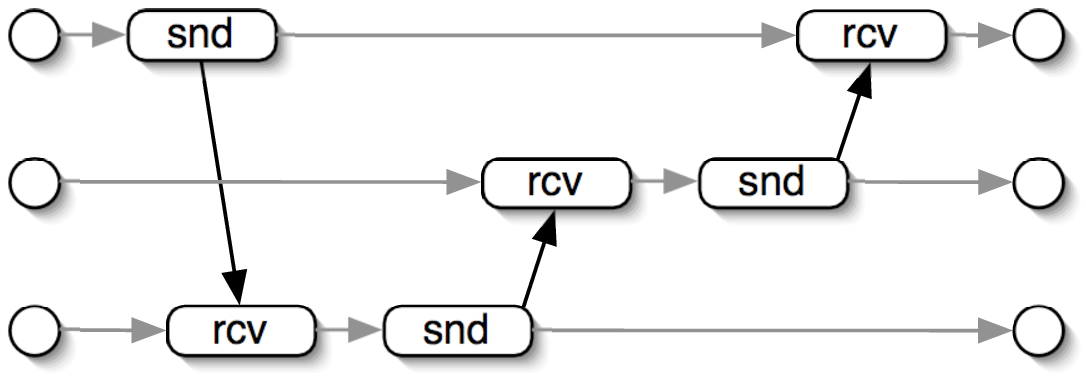}}
  \end{center}
  \caption{An example of a program for two processes that is
    unsealable unless a third process is added.}
  \label{fig:PGs}
\end{figure}

The above discussion suggests that it is necessary to inspect the next
layer in order to determine whether a given layer is a CCL. In fact, we
shall define a notion of a program $Q$ \emph{sealing} its predecessor
$P$, which will ensure that $P$ is a CCL in $P$ immediately followed by
$Q$.
For example, $\valt[j\to i]{}$ seals $\valt{}$ and vice versa.
Intuitively, $Q$ seals $P$ if $Q$ guarantees that no message sent
after $P$ can be received in $P$. Let us consider why $\valt[j\to
i]{\ack}$ seals $\valt{}$. Suppose that a later message is sent on the
channel from $i$ to $j$ as in Fig.~\ref{fig:vtseals}. This send is
performed only after the message sent in the opposite direction has
been received by $i$, which in turn must have been sent after the
first message has been received by $j$. Consequently, $j$'s receive
event must precede $i$'s sending of the later message. Therefore, the
later message cannot compete with the earlier one. A message
transmitted in the opposite direction is often called an
\emph{acknowledgment}.
More interesting examples of sealing are presented in
Figures~\ref{fig:3sealP} and~\ref{fig:PGs}.
For a decomposition of a program $P$ into a sequence of $\ell$ layers
$L^{(1)}\dotty L^{(\ell)}$, it follows that if $L^{(k+1)}$ seals
$L^{(k)}$ for all $1\leq k<\ell$ then each layer $L^{(k)}$ is a CCL in
$P$.

In~\cite{Lamport1978:cacm} Lamport defined causality among events of
asynchronous message passing systems. Causality implies temporal
precedence. As discussed above, transmitting an acknowledgment
guarantees that the receive of the first message causally precedes any
later sends on the same channel. Observe that the same effect could be
obtained by other means ensuring the intended precedence. For
instance, a causal chain consisting of a sequence of messages starting
at $j$, going through a number of intermediate processes, and ending
at~$i$ could be used just as well. While this \emph{transitive} form
of acknowledgment appears to be inefficient, a given message can play
a role in a number of transitive acknowledgments.
Fig.~\ref{fig:3sealP} illustrates a program consisting of the
transmission of three messages over three different channels. It is
sealed using transitive acknowledgments by the program displayed in
Fig.~\ref{fig:3sealS}, which sends only two messages.

Indeed, we shall later show how $O(n)$ messages can usefully
substitute for $\Omega (n^2)$ acknowledgments.
Not all programs can be sealed. We shall later prove that program $X$
shown in Fig.~\ref{fig:unseal} is unsealable. The same program
executed in the presence of a third process as in
Fig.~\ref{fig:unsealP} is, however, sealable. Any seal of this program
will necessarily use transitive acknowledgments as discussed above.
See Fig.~\ref{fig:unsealS} for an illustration of one way this program
can be sealed.


\paragraph{Contributions.}

The first main contribution of this paper is in the presentation of a
framework studying safe composition of layers of distributed programs
in different models of communication. Within the framework we define
notions including CCL and barriers. Moreover, it is possible to define
new notions such as sealing that play an important role in ensuring
safe composition. In this paper the power of the framework is
illustrated by a comprehensive study of safe composition in \rel. In a
companion paper~\cite{EM2005b} the framework is used to define
additional notions that are used to study safe composition in
FIFO-models with duplicating and/or lossy channels.

Our second main contribution is in identifying the notion of sealing
and demonstrating its central role in the design of CCLs in \rel. We
study the theory of sealing in \rel and present the following results.
\begin{itemize}
\item Sealable straight-line programs are completely characterized.
\item A definition of the sealing \emph{signature} of straight-line
  programs is given, which characterizes the sealing behavior of a
  program concisely, for both purposes, sealing and being sealed.
  The size of the signature is $O (n^2)$.
\item An algorithm for deciding whether $Q$ seals $P$ based only on
  their signatures is presented.
\item An algorithm for constructing seals for sealable straight-line
  programs is presented.  It produces seals that perform less than
  $3n$ message transmissions even though $\Omega(n^2)$ channels may
  need to be sealed.
\end{itemize}
The restriction to straight-line programs is motivated by the
undecidability of the corresponding problems for general programs.
Specifically, the halting problem can be reduced to each of these
problems for general programs.
As far as communication closure is concerned, straight-line programs
already display most of the interesting aspects relevant to the
subject of sealing.

\section{A Model of Distributed Programs with Layering}
\label{sec:prog}


In this section we define a simple language for writing
message-passing concurrent programs. Its composition operator
``$\lay$'' is called \emph{layering}. Layering subsumes the two more
traditional operators ``$;$'' and ``$\parallel$'' (as discussed by
Janssen in~\cite{Janssen1994:PhD}). The meaning of $P\lay Q$ is that
each process $i$ first executes its share of $P$ and then proceeds
directly to execute its share of $Q$. In particular, layering does not
impose any barrier synchronization between $P$ and $Q$. In other
words, in $P\lay Q$ process $i$ need not wait for any other processes
to finish their shares of $P$ before moving on to $Q$.
Consequently, programs execute between \emph{cuts} rather than global
states. We shall define a notion $r[c,d]\support P$ of a program $P$
\emph{occurring} over an \emph{interval} $r[c,d]$ between the cuts $c$
and $d$ of a run $r$.

Our later analysis will be concerned with CCLs $P$. Thus we need to
ensure that no message crosses any initial or final cut of an interval
over which $P$ occurs. A concise way of capturing this formally is via
a new language construct, the \emph{\chopname}, $\chop$. Writing
$\chop$ specifies that all communication channels are empty at this
cut. In other words, no statement to the left of the $\chop$ can
communicate with a statement to the right. If $P$ is a CCL in a given
larger program $L$ then every execution of $P$ in $L$ is also an
execution of $\chop P\chop$. In other words, $P$ can be substituted
for $\chop P\chop$ in $L$.\footnote{In place of the \chopname $\chop$
  the preliminary version of this paper~\cite{EM2005a} used a
  \emph{phase quantifier} $\tau$. Program $\tau P$ roughly corresponds
  to our $\chop P\chop$.}
We adopt a standard notion of \emph{refinement} to indicate
substitutability of programs. Program $P$ \emph{refines} program $Q$
if every execution of $P$ over an interval $r[c,d]$ is also one of
$Q$, regardless of what happens before $c$ and after $d$.
The notions of ``$\lay$'', ``$\chop$'', and refinement provide a
unified language for defining notions of safe composition. The
programming language and its semantics are formally defined as
follows.

\subsection{Syntax}

Let $n\in\nat$ and $\agents = \{1\dotty n\}$ be a set of processes.
Throughout the paper $n$ will be reserved for denoting the number of
processes. Let $(\Var_i)_{i\in\agents}$ be mutually disjoint sets of
\emph{program variables (of process $i$)} not containing the name
$h_i$ which is reserved for $i$'s \emph{communication history}.
Let $\Expr_i$ be the set of arithmetic expressions over $\Var_i$.
Let $\Lan$ be propositional logic over atoms formed from expressions
with equality ``$=$'' and less-than ``$<$''.
We define a syntactic category $\Prog$ of \emph{programs}:
\begin{displaymath}
  \Prog \ni P \deqBNF \EmptyProg
  \altBNF x \Ass \expr
  \altBNF \snd[i\to j]{\expr}
  \altBNF \rcv[j\from i]{x}
  \altBNF \guard{\phi}
  \altBNF \chop
  \altBNF P \lay P
  \altBNF P \cho P
  \altBNF P\rpt
\end{displaymath}
where $x\in\Var_i$, $\expr\in\Expr_i$, $i,j\in\agents$, and $\phi \in \Lan$.

The intuitive meaning of these constructs is as follows.  The symbol
$\EmptyProg$ denotes the \emph{empty program}. It takes no time to
execute. \emph{Assignment statement} $x\Ass \expr$ evaluates expression
$\expr$ and assigns its value to variable $x$. The $\snd[i\to j]{\expr}$
statement sends a message containing the value of $\expr$ on the channel
from $i$ to $j$. Communication is asynchronous, and sending is
non-blocking. The $\rcv[j\from i]{x}$ statement, however, blocks until a
message arrives on the channel from $i$ to $j$. It takes a message off
the channel and assigns its content to $x$.
The \emph{guard} $\guard{\phi}$ expresses a constraint on the
execution of the program: in a run of the program, $\phi$ must hold at
this location. Guards take no time to execute. The program $\chop$ is
a guard-like constraint stating that all channels must be empty at
this location. Formally, our propositional language $\Lan$ is not
expressive enough to define $\chop$ as a guard because formulas are
not capable of refering to channel contents. The operation ``$\lay$''
represents \emph{layered composition} following Janssen et
al.~\cite{Janssen95:TACAS}. Layering statements of distinct processes
is essentially the same as parallel composition whereas layering of
statements of the same process corresponds to sequential composition.
We tend to omit ``$\lay$'' when no confusion will arise. The symbol
``$\cho$'' denotes 
nondeterministic choice. By $P\rpt$ we denote zero or more (possibly
infinitely many) repetitions of program $P$.\footnote{Using guards,
  choices, and repetition it is possible to define
  \begin{math}
    \kw{if}\;\phi\;\kw{then}\;P\;\kw{else}\;Q\;\kw{fi}
  \end{math}
  as an abbreviation for
  \begin{math}
    \guard{\phi} \omitlay P \cho \guard{\Not\phi} \omitlay Q
  \end{math}
  and
  \begin{math}
    \kw{while}\;\phi\;\kw{do}\;P\;\kw{od}
  \end{math}
  for
  \begin{math}
    (\guard{\phi} \omitlay P)\rpt \omitlay\guard{\Not\phi}
  \end{math}. The results in this paper also hold for a language based
  on $\kw{if}$ and $\kw{while}$ instead of $\guard{.}$, $\cho$, and
  $\rpt$.}

\subsection{Semantics}

A \emph{send record (for $i$)} is a triple $(i\to j, v)$, which
records sending a message with contents $v$ from $i$ to the receiver
$j$. Similarly, $(j\from i, v)$ is a \emph{receive record (for $j$)}.
A \emph{local state (for process $i$)} is a mapping from $\Var_i$ to
values and from $h_i$ to a sequence of send and receive records for $i$.
A \emph{local run (for process $i$)} is an infinite sequence of local
states. We identify an \emph{event (of $i$)} with the transition from
one local state in a local run of $i$ to the next. An event is either
a \emph{send}, a \emph{receive}, or an \emph{internal} event. A
\emph{(global) run} is a tuple $r=((r_i)_{i\in\agents}, \delta_r)$ of
local runs --- one for each process --- plus an injective
\emph{matching function} $\delta_r$ associating a send event with each
receive event in $r$. The mapping $\delta_r$ is restricted such
that:\footnote{Our choice of execution model is closely related to the
  more standard one of infinite sequences of global states,
  representing an \emph{interleaving} of moves by processes. Our
  conditions on $\delta_r$ guarantee the existence of such an
  interleaving. In general, each of our runs represents an equivalence
  class of interleavings.}
\begin{enumerate}
\item If $\delta_r (e) = e'$ and $e$ is a receive event of process $j$
  resulting in the appending of $(j\from i,v)$ to $j$'s message
  history then $e'$ is a send event of process $i$ appending the
  corresponding send record $(i\to j,v)$ to $i$'s message history.
\item Lamport's causality relation $\lacause$ induced by $\delta_r$ on
  the events of $r$, as defined below, is an irreflexive partial
  order, hence acyclic.
\end{enumerate}
The first condition captures the property that messages are not
corrupted in transit. The fact that the function $\delta_r$ is total
precludes the reception of spurious messages, whereas injectivity
ensures that messages are not duplicated in transit. Further
restrictions on $\delta_r$ can be made to capture additional
properties of the communication medium such as reliability, FIFO,
fairness, etc.

We say that $r\in\rel$ if no unmatched send event is succeeded by
infinitely many matched send events on the same channel.

In~\cite{Lamport1978:cacm} Lamport defined a ``happened before''
relation $\lacause$ on the set of events occurring in a run $r$ of a
distributed system. The relation $\lacause$ is defined as the smallest
transitive relation subsuming (1) the total orders on the events of
process $i$ given by the $r_i$, and (2) the relation
$\setcpr{(e_1,e_2)}{\delta_r (e_2) = e_1}$ between send and receive
events induced by the matching function $\delta_r$.

\subsubsection{Cuts and Channels}

Write $\natplus$ for $\nat \cup \{\infty\}$. A \emph{cut} is a pair
$(r,c)$ consisting of a run $r$ and a $\agents$-indexed family $c =
(c_i)_{i\in\agents}$ of $\natplus$-elements.  We write ``$\leq$'' for
the component-wise extension of the natural ordering on $\natplus$ to
cuts within the same run. A cut is \emph{finite} if all its components
are.

Say that an event $e$ performed by process $i$ is \emph{in} a cut
$(r,c)$ if $e$ occurs in $r_i$ at an index no larger than~$c_i$,
and~$e$ occurs \emph{outside} of $(r,c)$ if the index is larger
than~$c_i$.
A cut $(r,c)$ corresponds to the, possibly implausible, situation in which
the events in the cut have occurred for each process $i\in\agents$. 
We define the \emph{channel $\chan{i}{j}$} at a cut $(r,c)$ to be the
set of $i$'s send events to $j$ and $j$'s receive events from $i$ in
$(r,c)$ that are not matched by $\delta_r$ to any event also in
$(r,c)$.
Finally, a formula $\phi\in\Lan$ \emph{holds at $(r,c)$}, and we write
$(r,c) \models \phi$, if $\phi$ holds in standard propositional logic
when, for each $i\in\agents$, program variables in $\Var_i$ are
evaluated in the local states $r_i(c_i)$ if $c_i$ is finite, and are
considered unspecified otherwise.\footnote{Recall that local states
  assign values to local variables.}

Observe that a cut can, in general, be fairly arbitrary. In
particular, there is no requirement that all messages that are
received before a cut is reached were sent before the cut. This is
deliberate. There are, of course, many instances in which more
structured cuts may be of interest. Indeed, we can define a cut
$(r,c)$ to be \emph{consistent} if every $\lacause$ predecessor of an
event in the cut $(r,c)$ is also in the cut. Moreover, in this work we
make use of a stronger property of cuts---that all channels are
\emph{empty} at the cut.

\subsubsection{Semantics of Programs}

We define the meaning of programs by stating when a program occurs
over an interval.
An \emph{interval} consists of two cuts $(r,c)$ and $(r,d)$ over the
same run with $c\leq d$, which we denote for simplicity by $r[c,d]$.
An event is \emph{in} $r[c,d]$ if it is in $(r,d)$ but not in $(r,c)$.
We define the occurrence relation $\support$ between intervals and
programs by induction on the structure of programs. The interesting
cases are those of $\lay$ and $\chop$. Formally, program $P\in\Prog$
\emph{occurs} over interval $r[c,d]$, denoted $r[c,d] \support P$,
iff:\footnote{We shall denote by $f[a\mapsto b]$ the function that
  agrees with $f$ on everything but $a$, and maps $a$ to $b$.}
\begin{trivlist}
\item $r[c,d] \support\EmptyProg$ if $c = d$.
\item $r[c,d] \support x\Ass \expr$ if $d = c[i\mapsto c_i+1]$ and
  $r_i(d_i) = r_i(c_i) [x\mapsto v]$, where $v$ is the value of $\expr$ in
  $r_i(c_i)$.
\item $r[c,d] \support\snd[i\to j]{\expr}$ if $d = c[i\mapsto c_i+1]$ and
  $r_i(d_i) = r_i(c_i) [h_i \mapsto r_i(c_i)(h_i)\cdot\langle (i\to j,
  v)\rangle]$, where $v$ is the value of $\expr$ in $r_i(c_i)$.
\item $r[c,d] \support \rcv[i\from j]{x}$ if $d = c[i\mapsto c_i+1]$
  and $r_i(d_i) = r_i(c_i) [h_i \mapsto r_i(c_i)(h_i) \cdot\langle (i\from
  j, v)\rangle, x \mapsto v]$.
\item $r[c,d] \support\guard{\phi}$ if $c=d$ and $(r,c) \models \phi$.
\item $r[c,d] \support\chop$ if $c=d$ and no
  communication
  event in $(r,c)$ is matched by $\delta_r$ with an event outside
  $(r,c)$.\footnote{I.e., no receive in the cut $(r,c)$ is mapped by
    $\delta_r$ to a send outside of the cut, and no receive from
    outside is mapped to a send in the cut.}
\item $r[c,d] \support P \lay Q$ if there exists $c'$ satisfying
  $c\leq c'\leq d$ such that $r[c,c'] \support P$ and $r[c',d]
  \support Q$.
\item $r[c,d] \support P \cho Q$ if $r[c,d] \support P$ or $r[c,d]
  \support Q$.
\item $r[c,d] \support P\rpt$ if, intuitively, an infinite or finite
  number (possibly zero) of iterations of $P$ occur over $r[c,d]$.
  More formally, $r[c,d] \support P\rpt$ if there exists a finite or
  infinite sequence $(c^{(k)})_{k\in I}$ such that $I$ is a non-void
  prefix of $\natplus$, $c^{(0)} = c$, $c^{(k)} \leq c^{(k')}$ for all
  $k<k'\in I$, $\lub_{k\in I} c^{(k)} = d$, and $r[c^{(k)},c^{(k+1)}]
  \support P$ for all $k,k+1\in I$.
\end{trivlist}
The program semantics is insensitive to deadlocks because deadlocking
executions are not represented by runs.
We deliberately chose to ignore deadlocks to simplify the presentation
and focus on the main aspects of composition. Whether a program
deadlocks can be analyzed using standard
techniques~\cite[p.~635f]{Lynch96}.

\paragraph{General assumption.}

\emph{From now onward, we shall only consider programs that are
  deadlock-free
  .}

\subsubsection{Refinement}

We shall capture various assumptions about properties of systems by
specifying sets of runs. For instance, \rel is the class of runs with
reliable communication, and \relfifo is its subclass in which channels
are also FIFO.

Given a set $\Gamma$ of runs, we say that \emph{$P$ refines $Q$ in
  $\Gamma$}, denoted $P \refines_\Gamma Q$, iff $r[c,d] \support P$
implies $r[c,d] \support Q$, for all $r\in\Gamma$ and
$c,d\in(\natplus)^\agents$. In other words, every execution of $P$ (in
a $\Gamma$ run) is also one of $Q$, regardless of what happens before
and after.  Therefore, we may replace $Q$ by $P$ in any larger program
context. 
This definition of refinement is thus appropriate for stepwise
top-down development of programs from specifications.
The refinement relation on programs is transitive (in fact a
pre-order) and all programming constructs are monotone w.r.t.~the
refinement order.


\section{Capturing Safe Composition}

The \chopname program $\chop$ allows us to delineate the interactions
that a layer can have with other parts of the program. When combined
with refinement it is useful for defining various notions central to
the study of safe composition, as we now illustrate.

\paragraph{CCL.}
We can express that the program $L$ is a CCL in the program $P\lay
L\lay Q$ w.r.t.~$\Gamma$ by:
\begin{gather*}
  \chop P \lay L \lay Q \chop \quad\refines_\Gamma\quad
  P \chop L \chop Q\enskip\text{.}
\end{gather*}
In words, any isolated execution of $P\lay L\lay Q$ will have the
property that all communication in $L$ is internal and hence $L$
executes as in isolation. This definition is context-sensitive.

\paragraph{Barriers.}

More modular would be a notion that guarantees safe composition
regardless of the program context. One technique to ensure that two
consecutive layers do not interfere with each other is to place a
barrier $B$ between them.
Formally, program $B$ is a \emph{barrier} in~$\Gamma$
if
\begin{gather*}
  \chop P\lay B\lay Q \quad\refines_\Gamma\quad P\chop B \chop
  Q\enskip\text{, for all $P,Q
    $.}
\end{gather*}
Traditionally, barriers have been used to synchronize the progression
through phases by enforcing that no process could start its
$n+1^\text{st}$ task before all the other processes had completed
their $n^\text{th}$ tasks.
This could be formilzed by requiring that, if $r[c,c']\support \chop
P$, $r[c',d']\support B$, and $r[d',d]\support Q$, then all events in
$(r,c')$ necessarily $\lacause$-precede all events not in $(r,d')$,
for all runs $r\in\Gamma$, and programs $P,Q
$.

\paragraph{TCC.}

Some programs can be safely composed without the need for
communication-closedness~\cite{Elrad.Francez82:scp,Janssen.Zwiers:92}.
Depending on the model $\Gamma$, there may be programs $P$ that safely
compose with all following layers. We say that \emph{$P$ is \tcc
  (\TCC)} in $\Gamma$ if,
\begin{gather*}
  \chop P
  \quad\refines_\Gamma\quad
  P \chop
  \enskip\text{.}
\end{gather*}
Thus, if $P$ is \TCC then any execution of $P$ starting in empty
channels will also end with all channels empty. Therefore \TCC
programs can be readily composed.\footnote{\TCC follows and is closely
  related to the notion of GTCC introduced by Gerth and
  Shrira~\cite{GerthShrira1986:FSTTCS}. The main difference is that
  their notion is defined w.r.t.~a set of initial states.}
It is straightforward to check that the programs $\EmptyProg$,
$\guard{\phi}$, $x\Ass \expr$, and $P\chop$ are \TCC in any
$\Gamma$. Moreover, if $P$ and $Q$ are \TCC in $\Gamma$ then so are
$P\cho Q$, $P\lay Q$, and $P\rpt$.

Observe that every barrier $B$ in $\Gamma$ is in particular \TCC in
$\Gamma$.

\paragraph{Seals.}

In many models of interest, only trivial programs are TCC\@. This is the
case, for example, in \rel, as shown in Section~\ref{sec:rel} below.
In such models, an alternative methodology is required for determining
when it is safe to compose given programs. Next we define a notion of
\emph{sealing} that formalizes the concept of program $S$ serving as
an impermeable layer between $P$ and later phases such that no later
communication will interact with $P$. We say that \emph{$S$ seals $P$
  in $\Gamma$} if,
\begin{displaymath}
  \chop P \lay S
  \quad\refines_\Gamma\quad
  P \chop S
  \enskip\text{.}
\end{displaymath}
Thus, if $S$ seals $P$ in $\Gamma$ then neither $S$ nor any later
program can interfere with communication in $P$. If $S$ seals $P$ and
$Q$ seals $S$, then $S$ will behave in $\chop P \lay S \lay Q$ as it
does in isolation.
Sealing allows incremental program development while maintaining
CCL-style composition.
\begin{lemma}\label{l:seal.algebra}
  \begin{enumerate}
  \item If both $P$ and $P'$ are sealed 
    by $S$ in $\Gamma$ then so is $P\cho P'$.
  \item If both $S$ and $S'$ 
    seal $P$ in $\Gamma$ then $S \cho S'$ (properly) seals $P$ in
    $\Gamma$.
  \item If $S$ seals $P$ in $\Gamma$ then $S\lay Q$ seals $P$ in
    $\Gamma$.
  \item If both $S$ 
    seals $P$ and $S'$ 
    seals $S$ in $\Gamma$, then $S'$ 
    seals $P\lay S$ in $\Gamma$.
  \item If $P$ 
    seals itself in $\Gamma$ then $P$ 
    seals $P\rpt$ in $\Gamma$.\label{it:l:seal.algebra.rpt}
  \item TCC subsumes sealing: $P$ is \TCC in $\Gamma$ iff all programs
    seal $P$ in $\Gamma$.
  \end{enumerate}
\end{lemma}
It follows from this lemma that, if program $P$ can be decomposed into
a sequence of $\ell$ layers $L^{(1)}\dotty L^{(\ell)}$, and in
addition $L^{(k+1)}$ seals $L^{(k)}$ for all $1\leq k<\ell$, then each
layer $L^{(k)}$ is a CCL in $P$.

For example, as discussed in the introduction, any program of the form
$\valt[j\to i]{}$ seals any program of the form $\valt{}$ in \rel.
Consequently, a program of the form $\valt{} \lay \valt[j\to i]{}$
seals itself in \rel. On the other hand, the shorter program $\valt{}$
does not seal itself in \rel---in an execution of $\valt{}\lay\valt{}$
the two messages sent by $i$ could be received in the reverse order of
sending.

\paragraph{Proper Seals.}

Suppose that $\agents = \{1,2\}$ and $x_i\in\Var_i$ for $i\in\agents$.
Then the program $Q = \kw{while}\;\True\;\kw{do}\;(x_1\Ass 5\lay
x_2\Ass 17)\;\kw{od}$ is \TCC in \relfifo, a CCL in \rel, and seals
any program in \rel. For it necessarily \emph{diverges}, that is, it
occurs only over intervals $r[c,d]$ with non-finite $d$. This implies
that no layer following $Q$ has any impact on the semantics of the
whole program. It follows trivially that no communication of a later
layer can interfere with anything before. Programs such as $Q$ are not
particularly useful as seals, in contrast to ones that seal without
diverging.
This motivates the following definition.
We say that $S$ is a \emph{proper seal} of $P$ in $\Gamma$ if $S$
seals $P$ and $S$ never diverges after $P$. That is, for all
$r\in\Gamma$ and $c,d,d'$, whenever $r[c,d] \support \chop P$, and
$r[d,d'] \support \chop S$ and $d$ is finite then so is $d'$.

For instance, since $\valt{}$ is a terminating program that seals
$\valt[j\to i]{}$ in \rel, it is in particular a proper seal.

\section{Case Study: Safe Composition in \rel}
\label{sec:rel}

We now consider safe composition in the model \rel. Communication
events can cause a program \emph{not} to be \TCC in \rel. For example,
reconsider the program $\valt{\expr\to x} = \snd[i\to
j]{\expr}\lay\rcv[j\from i]{x}$. It is \TCC in \relfifo but not \TCC
in \rel. That $\valt{}$ is not \TCC in \rel is no coincidence. Next we
show that no terminating program performing any communication
whatsoever is TCC in \rel.
\begin{theorem}\label{th:TCCbadforREL}
  If $r[c,d]\support P$ for some $r\in\rel$ and finite $c,d$ such that
  all channels are empty in $(r,c)$ and there is at least one send or
  receive event in $r[c,d]$, then $P$ is not \TCC in \rel.
\end{theorem}
\begin{proof}
  Assume that $r[c,d]\support P$ where $r\in\rel$, $c,d$ are finite,
  all channels are empty at $(r,c)$ and there is a send or receive
  event in $r[c,d]$.  If there is a non-empty channel in $(r,d)$ the
  claim is immediate since a matching communication event following
  $P$ could interact with $P$. Otherwise, every message sent in
  $r[c,d]$ is received in $r[c,d]$. Since $P$ is deadlock-free by the
  general assumption, there are processes whose last communication
  event in $r[c,d]$ is a receive. W.l.o.g.~let $i$ be such a process
  and assume that its last receive is of a message $v$ sent by $j$
  into variable $x\in\Var_i$.
  
  A run $r'\in\rel$ that equals $r$ up to $d$ can be constructed such
  that $r'[d,d'] \support \chop \snd[j\to i]{\expr} \lay \rcv[i\from
  j]{x} \chop$, where $\expr$ evaluates to $v$ in $r_j(d_j)$. So the
  same message is transmitted twice between $j$ and $i$. Let
  $r''\in\rel$ be the same as $r'$, except for $\delta_r$, which swaps
  the matching send events between the two receive events. For $Q =
  \snd[j\to i]{\expr} \lay \rcv[i\from j]{x}$ it follows that
  $r''[c,d'] \support \chop P\lay Q \chop$ but $r''[c,d'] \not\support
  \chop P \chop Q \chop$. The claim follows.
  \qed
\end{proof}
Since a barrier is necessarily \TCC we immediately obtain
\begin{corollary}
  No program can serve as a barrier in \rel
.
\end{corollary}
Having shown that \TCC and thus barriers are not generally useful
notions in \rel, we turn our attention to (proper) sealing. It is
instructive that not all terminating programs can be properly sealed
in \rel:
\begin{lemma}\label{l:non-sealed}
  If $\agents = \{1,2\}$ then the program $X = \snd[1\to 2]{x+1}\lay
  \snd[2\to 1]{y+1}\lay\rcv[1\from 2]{x}\lay\rcv[2\from 1]{y}$
  illustrated in Fig.~\ref{fig:unseal} cannot be sealed properly in
  \rel.
\end{lemma}
\begin{proof}
  Assume, by way of contradiction, that $S$ properly seals $X$ in \rel.
  Consider a run $r\in\rel$ such that $r[(0,0),(2,2)] \support X$ and
  $r[(2,2), d']\support S$ where $d'$ is finite.  If some process
  $i\in\agents$ does not engage in any communication event in
  $r[(2,2), d']$ then $S$ does not seal $X$ since a send by process
  $i$ performed at $d'_i$ potentially interacts with $X$.  Otherwise,
  let $e_i$ be the first communication events of each process $i=1,2$
  in $r[(2,2), d']$. If one of the $e_i$ is a send then, as before,
  this send can interact with $X$. Finally, if both $e_i$ are receives
  then $S$ causes a deadlock, contradicting the assumption that
  $r[(2,2), d']\support S$.
  \qed
\end{proof}
Our programming language $\Prog$ is Turing-complete. Since the halting
problem for $\Prog$ can be reduced to sealability in \rel we obtain
\begin{theorem}
  Sealability in \rel is undecidable.
\end{theorem}
Given this theorem we shall restrict our attention to more tractable
subclasses of programs.
Program $P$ is \emph{balanced (in \rel)} if, whenever $r[c,d]\support
P$ and all channels are empty at $(r,c)$, then every channel contains
an equal number of sends and receives at $(r,d)$. Note that balanced
programs are TCC in \relfifo. The following theorem shows that in \rel
balance is a necessary prerequisite for being properly sealable.
\begin{theorem}\label{th:psealable.balanced}
  In \rel, every non-divergent program that is properly sealable is
  also balanced.
\end{theorem}
\begin{proof}
  \newcommand{\ms}{k}
  \newcommand{\mr}{m}
  Let $P$ and $S$ be programs such that in \rel $P$ does not diverge
  and $S$ properly seals $P$.  Assume by way of contradiction that $P$
  is not balanced.  Let $r\in\rel$ and $c,c',d$ be such that
  $r[c,c']\support P$, $r[c',d]\support S$, all channels are empty in
  $(r,c)$, and, w.l.o.g., $\chan{i}{j}$ contains $\ms$ sends and
  $\mr$ receives at $(r,c')$ where $\ms\neq \mr$.  Since $S$
  is a proper seal, there is neither a send nor a receive event in
  $r[c,c']$ matched with an event not in $r[c,c']$. Since every
  receive event must be matched to some event by $\delta_r$ it follows
  that $\ms > \mr$, that is, there are more sends than receives
  on $\chan{i}{j}$ in $r[c,c']$. No receive in the seal can be matched
  to any of those sends. There exist $r'\in\rel$, $y\in\Var_j$, and
  $d'$ such that $r'$ is the same as $r$ up to $d$ (hence
  $r'[c,d]\support P\lay S$), $r'[d,d'] \support \rcv{j}$, and
  $\delta_r$ maps the receive event $r_j(d_j)$ to one of the send
  events of $P$ that are unmatched in $r$. This match contradicts the
  assumption that $S$ properly seals $P$.
  \qed
\end{proof}
Program $P$ is said to \emph{close $\chan{i}{j}$ (in \rel)} if
$\chan{i}{j}$ is empty after $P$ in any execution of $P$ starting at a
cut with empty channels. More formally this is expressed as follows.
For all $r\in\rel$, if $r[c,d]\support \chop P$ then $\chan{i}{j}$ is
empty in $(r,d)$. A channel that is not closed is \emph{open}. The
state of a program's channels is the essential element in determining
sealability.

Program $P$ is \emph{straight-line} if it contains neither
nondeterministic choices nor loops nor guards. In other words, $P$ is
built from sends, receives, and assignments using layering only. Our
focus in this section is on balanced straight-line programs, or
\emph{\BSL} for short.

The \emph{program graph} of a \BSL $P$ is a graph $(V,E)$ that has a
node for every send and receive event in $P$ plus an initial dummy
node $\Fd_i$ and a final dummy node $\Ld_i$ for each process $i$. The
edge set $E$ consists of the successor relation over events in the
same process extended to the dummy nodes plus an edge between the
$k$'th send and the $k$'th receive on channel $\chan{i}{j}$, for all
$k$, $i$, and $j$. All the graphs in Figures~\ref{fig:3seal}
and~\ref{fig:PGs} are program graphs. The size of a \BSL $P$'s program
graph is of the order of the size of the program.

Next we investigate the connection between program graphs and Lamport
causality. We use~$E^+$ to refer to the irreflexive transitive closure
of~$E$ and call edges not containing dummy nodes \emph{normal}. The
subset of normal edges is denoted by $N_E$. In \relfifo, the normal
edges induce the full causality relation on the events of the program.
As we shall show, in \rel the normal edges of a program graph are also
$\lacause$ edges.
\begin{lemma}
  \label{GraphCause}
  Let $r\in\rel$. Let $P$ be a \BSL with program graph $(V,E)$. If
  $r[c,d] \support \chop P$ then $N_E \subseteq \mathord{\lacause}$.
\end{lemma}
\begin{proof}
  The only interesting normal edges are those between sends and
  receives of different processes. Consider the edge $(e_1,e_2)\in E$
  between the $k$'th send and the $k$'th receive on $\chan{i}{j}$. Let
  $r\in\rel$ such that $r[c,d] \support \chop P$ and assume that $e_3
  = \delta_r (e_2)$ is the $\ell$'th send on $\chan{i}{j}$ in $P$. We
  need to show that $e_1 \lacause e_2$. By definition of $\lacause$,
  we have that $e_3 \lacause e_2$. If $\ell = k$ then $e_3 = e_1$ and
  we are done. If $\ell > k$ then $e_1 \lacause e_3$ because $e_1$ is
  an earlier event of $i$ than $e_3$ and the claim follows by
  transitivity of $\lacause$. Finally, suppose that $\ell < k$. This
  case is illustrated in Fig.~\ref{fig:lproof}.
  \begin{figure}[htbp]
    \centering
    \includegraphics[width=8.5\grafflecm]{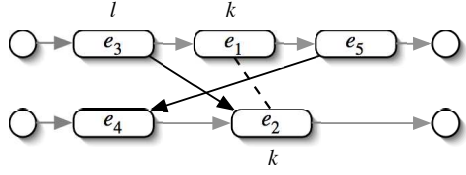}
    \caption{The case $\ell < k$ in the proof of Lemma~\ref{GraphCause}.}
    \label{fig:lproof}
  \end{figure}
  Consider the $k-1$ receives on $\chan{i}{j}$ that precede~$e_2$.
  They are all matched in~$r$ to sends by~$i$. Since $\ell<k$ and
  $e_3$ is already matched to~$e_2$, one of these receives, say $e_4$,
  must be matched to a send event~$e_5$ that does not precede~$e_1$.
  Since $e_5\lacause e_4$ and $e_4\lacause e_2$, it follows that
  $e_1\lacause e_2$, as desired.
  \qed
\end{proof}
Lemma~\ref{GraphCause} implies that all edges in $(N_E)^+$ will be
$\lacause$ edges in every run~$r\in\rel$ of~$\chop P\lay Q$. We
note that $(N_E)^+$ is the largest set of edges with this property,
because $(N_E)^+ = (\mathord{\lacause}\cap V^2)$ if $r\in\relfifo$.

A more concise representation than the program graph is called the
\emph{signature of $P$} and denoted by $\sig (P)$. It has size
$O(\cardP^2)$ while preserving the information necessary to decide
what channels are left open, respectively closed, by $P$. Given the
program graph $(V,E)$ of a \BSL $P$ we can obtain $\sig (P)$ as
follows. After calculating $E^+$, we remove all nodes except for the
dummy nodes and the first send and last receive on each channel. The
graph is further reduced by removing the node $\snd{}$ whenever
$(\Fd_j, \snd{})\in E^+$. Similarly, $\rcv{}$ is removed whenever
$(\rcv{}, \Ld_i)\in E^+$. The sends and receives remaining in the
signature are precisely the ones that could interfere with receives in
a preceding layer or with sends in a succeeding layer.

The complexity of computing $\sig (P)$ is in $O(\size{P}^3)$
since it requires the causality relation obtained as the transitive
closure of the edge relation of $P$'s program graph. We remark that
for \BSLs $P$ and $Q$, $\sig (P\lay Q)$ can be obtained from
their respective signatures at a cost of $O (\cardP^2)$.
 
\begin{figure}[thbp]
  \begin{center}
    \subfigure[$\sig(\valt{})$
    \label{fig:Sigvt}]{%
      \includegraphics[width=4.5\grafflecm]{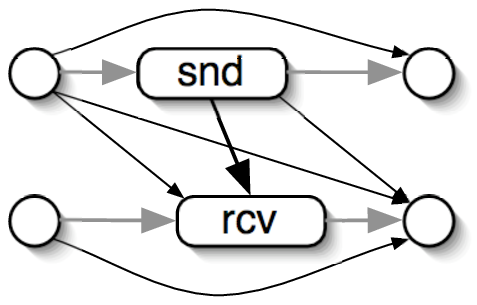}}\goodgap
    \subfigure[$\sig(\valt{}\lay\valtji)$
    \label{fig:Sigvtack}]{%
      \includegraphics[width=7.1\grafflecm]{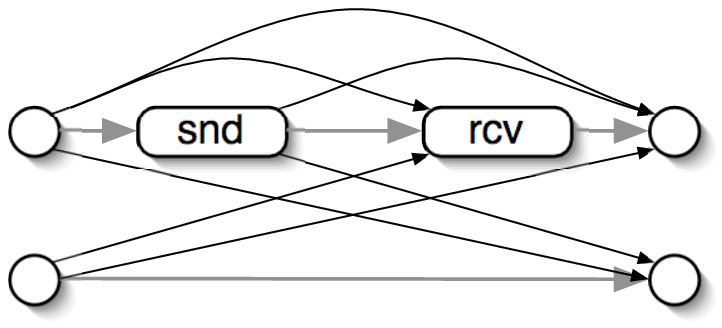}}
  \end{center}
  \caption{Examples of signatures. Thin arrows denote transitive
    causality edges.}
  \label{fig:Sigs}
\end{figure}

Let $P$ be a \BSL and let $G = (V,E)$ be $\sig (P)$. Then $P$ leaves
channel $\chan{i}{j}$ open iff $\rcv[j\from i]{} \in V$%
. For instance, the program $\valt{}$ leaves $\chan{i}{j}$ open ---
there is a node $\rcv{}$ in $\sig (\valt{})$, which is depicted in
Fig.~\ref{fig:Sigvt}. As we have shown earlier, $\valtji$ seals
$\valt{}$ in \rel, which implies that $\valtji$ closes $\chan{i}{j}$
once. Since $\valtji$ does not re-open the channel, the $\rcv{}$ node
found in $\sig(\valt{})$ is not present in the $\sig
(\valt{}\lay\valtji)$ shown in Fig.~\ref{fig:Sigvtack}.

\subsection{Deciding Sealing}

\begin{figure}[thbp]
  \begin{center}
    \subfigure[Channel $\chan{i}{j}$ left open by $P$ and a causality
    edge to $\Ld_k$.
    \label{fig:SigPcharSeal}]{%
      \goodgap
      \includegraphics[width=5.5\grafflecm]{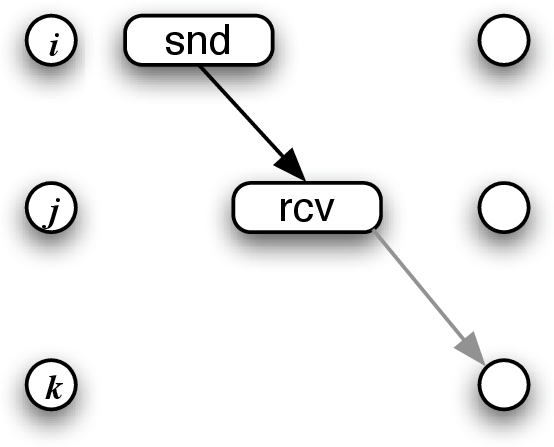}\goodgap}\goodgap
    \subfigure[Sealing the channel by causality. The dashed part
    accounts for $Q$ sending on $\chan{i}{j}$.
    \label{fig:SigQcharSeal}]{%
      \goodgap
      \includegraphics[width=5.5\grafflecm]{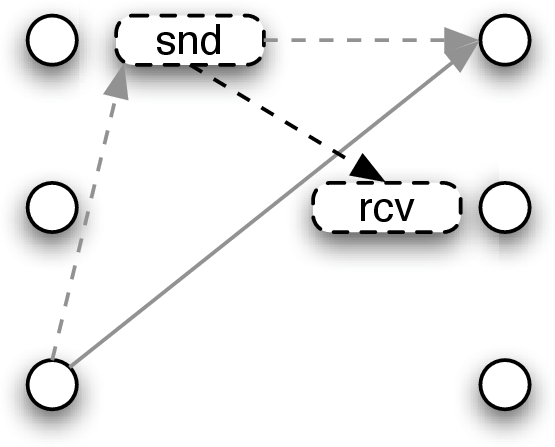}\goodgap}
  \end{center}
  \caption{Excerpts of the signatures of \BSLs $P$ and $Q$.}
  \label{fig:charseal}
\end{figure}

Whether one \BSL seals another can be decided on the basis of their
signatures. Suppose \BSL $P$ leaves $\chan{i}{j}$ open and $Q$ seals
$P$. Then, if $Q$ sends on that channel, then $P$'s last receive
$\rcv{}$ on the channel must causally precede $Q$'s first send
$\snd{}$ on it. Otherwise, $Q$ must ensure that any later send on
$\chan{i}{j}$ is causally preceded by $P$'s last receive. This is
guaranteed exactly if $P$'s signature contains an edge
$(\rcv{},\Ld_k)$ and $Q$'s signature contains an edge $(\Fd_k,
\Ld_i)$, for some $k\in\agents$. (See Fig.~\ref{fig:charseal}.)

Based on the above observation the following theorem
characterizes sealing among \BSLs.
\begin{theorem}\label{th:decide.sealing}
  Let $P$ and $Q$ be \BSLs and let $(V_P,E_P) = \sig (P)$ and
  $(V_Q,E_Q) = \sig (Q)$. Then $Q$ properly seals $P$ iff, for all
  $\rcv{}\in V_P$, there exists $k\in\agents$ such that
  $(\rcv{},\Ld_k)\in E_P$, $(\Fd_k, \Ld_i)\in E_Q$, and, if $\snd{}\in
  V_Q$ then $(\Fd_k, \snd{})\in E_Q$.
\end{theorem}
\begin{proof}
  \textbf{``$\ISIMPLIED$''} Consider the channel $\chan{i}{j}$.
  By construction, there is a node $\rcv{}\in V_P$ precisely if the
  channel is not closed by $P$. Suppose that $(\rcv{},\Ld_k)\in E_P$
  and $(\Fd_k, e)\in E_Q$ where $e = \snd{}$ if $\snd{}\in V_Q$ and
  $e=\Ld_i$ otherwise.  Let $r\in\rel$ and $c,d$ be such that $r[c,d]
  \support \chop P\lay Q$. Let
  $c'$ be such that $r[c,c'] \support P$ and $r[c',d] \support
  Q$. Let $\thercv$ in $r[c,c']$ be a $\rcv{}$ event. We shall prove
  that $\thesnd = \delta_r(\thercv)$ is also in $r[c,c']$. By
  definition of $\lacause$ we have that $\thesnd \lacause \thercv$.
  First observe that $\thesnd$ cannot be in $(r,c)$ since $r[c,d]
  \support \chop P\lay Q$ implies that $\delta_r$ cannot
  map $\thercv$ to an event in $(r,c)$. Second, $\thesnd$ cannot
  come after $(r,c')$ because, as we shall show, that would imply
  $\thercv \lacause \thesnd$. By transitivity, we would obtain
  $\thercv \lacause \thercv$, contradicting the irreflexivity of
  $\lacause$. Assume by way of contradiction that $\thesnd$ is not in
  $(r,c')$. If $\thesnd$ is in $r[c',d]$, that is, generated by $Q$,
  then $\snd{}\in V_Q$ represents a send event $\thesnd'$. This event
  is causally preceded by $\thercv$ because $(\thercv, \Ld_k)\in E_P$,
  $(\Fd_k, \snd{})\in E_Q$, and $\thesnd = \thesnd'$ or $\thesnd'
  \lacause \thesnd$. Otherwise, that is, if $\thesnd$ is not in
  $(r,d)$, it is causally preceded by $\thercv$ because $(\thercv,
  \Ld_k)\in E_P$, $(\Fd_k, \Ld_i)\in E_Q$, and $\thesnd$ is causally
  preceded by the last event of process $i$ in $r[c,d]$. In either
  case, $\thercv \lacause \thesnd$ follows by transitivity.
 
  By now we have shown that $\delta_r$ does not match any receive in
  $r[c,c']$ to a send event not in $r[c,c']$. Because $P$ is balanced
  this implies that all send events in $r[c,c']$ (i.e., the ones
  generated by $P$) must be matched with receive events in that
  interval. Thus, also $r[c,d] \support \chop P \chop Q$.
\item[]\textbf{``$\IMPLIES$''} Suppose that $\rcv{}\in V_P$ and that
  there is no $k$ such that $(\rcv{},\Fd_k)\in E_P$ and $(\Fd_k, e)\in
  E_Q$ where $e = \snd{}$ if $\snd{}\in V_Q$ and $e=\Ld_i$ otherwise.
  We show that $Q$ does not properly seal $P$. First consider the case
  $e = \snd{}$. For lack of a causal relationship between $\rcv{}\in
  V_P$ and $\snd{}\in V_Q$ they can be matched in an interval $r[c,d]$
  over which $\chop P\lay Q\chop$ occurs, violating the sealing
  property. Finally consider the remaining case, $e=\Ld_i$. Again, for
  lack of a causal relationship between $\rcv{}\in V_P$ and $\Ld_i\in
  V_Q$, a subsequent send event can be matched with $\rcv{}\in V_P$,
  that is, there exist $r\in\rel$ and $c,d$ such that $r[c,d]\support
  \chop P\lay Q \lay \snd{53}\chop$ and $\delta_r$ matches the last
  receive on channel $\chan{i}{j}$ in $P$ with the $\snd{53}$.
  \qed
\end{proof}
%
Given the theorem above, the complexity of deciding whether $Q$ seals
$P$, given their signatures, is obviously determined by the size of
$P$'s signature, which we recall is $O(\cardP^2)$.


\subsection{A Characterization of Sealability}

Observe that the set of channels closed by a \BSL $P$ when executed
from a cut with empty channels is uniquely determined by $P$ and can
be derived from its signature. We can thus associate a
\emph{closed-channel graph} with each \BSL\@. Formally, the
closed-channel graph $C_P = (\agents,E_P)$ of a \BSL $P$ is given by
$(i,j)\in E_P$ iff $i\neq j$ and $\chan{i}{j}$ is closed by $P$ in
\rel. In the following we denote the undirected version of a graph $G$
by $G\ud$.
\begin{theorem}[Sealability]
  \label{th:sealability}
  Let $P$ be a \BSL. Then $P$ can be sealed properly in \rel iff
  $C_P\ud$ is connected. Moreover, if $P$ is properly sealable in \rel
  then it can be sealed by a \BSL that transmits less than $3\cardP$
  messages.
\end{theorem}
\begin{proof}
  \textbf{``$\IMPLIES$''} Suppose that $C_P\ud$ is not connected.
  Then $\agents$ can be partitioned into two non-void sets, $A$ and
  $\comp{A}$, such that there is no channel closed by $P$ between
  (elements of) the two sets.  Assume, by way of contradiction, that
  the program $S$ properly seals $P$. Since $S$ is a seal, every
  message sent in $S$ along a channel not closed by $P$ must be
  causally preceded by all receives on that channel in $P$. This holds
  in particular for all channels between $A$ and $\comp{A}$. There
  must be such receives in $P$ for each of the channels not closed by
  $P$.  To establish the causal precedences, $S$ must transmit
  messages.  Unless $S$ transmits messages between $A$ and $\comp{A}$,
  it cannot seal $P$. Consider one of the causally minimal sends of
  such a transmission in $S$. It can interfere with the last receive
  on that channel in $P$.  Consequently, $S$ does not seal $P$.
\item\textbf{``$\ISIMPLIED$''} The algorithm sketched as
  $\proc{Seal}(P)$ takes a \BSL $P$ as input and outputs a proper seal
  for $P$ if $P$ is properly sealable.
  \begin{codebox}
    \Procname{$\proc{Seal}(P)$}
    \li $(V,E) \gets \proc{Closed-Channels}(P)$
    \>\>\>\>\>\>\>\>\>\>\Comment
    This algorithm is presented in Appendix~\ref{app:algo}.
    \li $S \gets \EmptyProg$    
    \li pick $T\subseteq E$ s.t.~$(\agents, T)\ud$ forms a spanning tree of $V$
    \li $v \gets$ a node at the center of $T$
    \li \For $(w,w')\in T$ pointing away from $v$ s.t.~$(w',w)\notin E$
    \label{li:Seal.For}
    \li     \Do
                $S \gets S\lay\valt[w\to w']{}$
                \label{li:Seal.Od}
            \End
    \li add a converge-cast in $T$ to $S$
    \li add a broadcast in $T$ to $S$
  \end{codebox}
  Let $S$ be the result of $\proc{Seal}(P)$. It consists of less than
  $3\cardP$ instances of $\valt[]{}$ because every spanning tree
  contains $\cardP-1$ edges and each of the three sub-phases, (a)
  lines~\ref{li:Seal.For}--\ref{li:Seal.Od}, (b) the converge-cast, and
  (c) the broadcast transmits less than $\cardP$ messages. Each one of
  these $\valt[]{}$ instances transmits a message along a channel that
  is closed at the time of transmission. For phase (a) this follows
  from the selection criterion for these transmission in
  line~\ref{li:Seal.For}. Phase (a) establishes that all channels
  between a node and its parent in the spanning tree are closed, thus
  phase (b) transmits on closed channels only. Similarly, phase (b)
  closes all channels between nodes and their children in the spanning
  tree, hence also phase (c) transmits on closed channels only.
  Finally, we need to show that every channel left open by $P$ is
  closed at least once by $S$. Let $(i,j)$ be such that $P$ leaves
  $\chan{i}{j}$ open. If $(i,j)\in T^{-1}$ then phase (a) closes the
  channel by sending on $\chan{j}{i}$. Otherwise it is closed
  transitively by the subsequence of the converge-cast from $j$ to the
  root $v$ followed by the subsequence of the broadcast from $v$ to
  $i$.
  \qed
\end{proof}
Observe that $\proc{Seal}(P)$ constructs a tailor-made tree barrier
$S$ between $P$ and any later program.
\begin{example}
  Consider a phase $L = \biglay_{i\in\agents}L_i$. In $L$ each process
  $i\neq 1$ sends a message to every other process $k\notin\{1,i\}$
  before receiving the $n-2$ messages sent to it in this phase.
  Finally, process $i$ transmits a message to process $1$. We can
  define process $i$'s program $L_i$ more formally by
  \begin{displaymath}
    L_i
    \quad=\quad
    \left(\biglay_{k\notin\{1,i\}} \snd[i\to k]{}   \right) \lay
    \left(\biglay_{k\notin\{1,i\}} \rcv[i\from k]{} \right) \lay
    \snd[i\to 1]{}\enskip\text{.}
  \end{displaymath}
  Process $1$ in turn receives those messages sent last in
  the $L_i$, that is:
  \begin{displaymath}
    L_1
    \quad=\quad
    \biglay_{i\neq 1} \rcv[1\from i]{}
  \end{displaymath}
  Executing $L$ beginning with empty
  channels leaves $n^2-3n+3$ channels open. Nevertheless, $L$ can be
  sealed efficiently by the program 
  \begin{displaymath}
    S
    \quad=\quad
    \biglay_{i\neq 1} (\snd[1\to i]{} \lay \rcv[i\from 1]{})\enskip\text{,}
  \end{displaymath}
  which transmits $n-1$ messages.
  (See Fig.~\ref{fig:maxtranssealS} for the program graph of $S$.)
  \begin{figure}[htbp]
    \begin{center}
      \includegraphics[width=8.5\grafflecm]{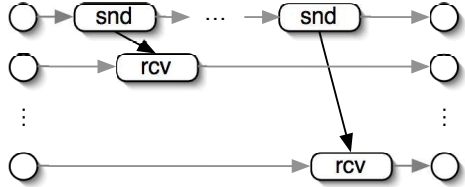}
    \end{center}
    \caption{$O(n)$ transmissions close $\Omega (n^2)$ open channels.}
    \label{fig:maxtranssealS}
  \end{figure}
\end{example}

\section{Conclusion and Future Work}
\label{sec:conclusion}

A subtle yet crucial issue in developing distributed applications is
the safe composition of smaller programs into larger ones. The notion
of CCL captures when a program works as if it were executed in
isolation in the context of a given larger program. The literature on
CCLs focused mostly on reliable FIFO communication. In that setting
programs can be designed that are inherently CCLs in any program
context.

Observe that neither termination detection nor barrier-style
techniques can be applied in \rel without careful inspection of the
surrounding program context. Any such mechanism will form a layer in
the resulting program which in turn must be shown to safely compose
with the other layers.
A popular approach to running distributed applications on non-\relfifo
systems is to construct an intermediate data-link layer providing
\relfifo communication to the application. This typically involves
sealing every single message transmission from interference by
previous and later layers. Popular algorithms for data-link achieve
this by adding message headers and/or acknowledging every single
message, thereby incurring a significant
overhead~\cite{A2F2LMWZ1994:jacm,WangZuck1989:PODC}. As we show for
\rel, it is often possible to do better than that. Our analysis of
sealing can be used to add the minimal amount of glue between
consecutive layers to ensure that they compose safely, without
changing the layers at all.

We have introduced a framework for studying safe program composition.
It facilitates the formal definition of standard notions such as CCL,
barriers, and TCC\@. Gerth and Shrira showed that---as a
context-sensitive notion---CCL is unsuitable for compositional
development of larger systems from off-the-shelf components. As we
have shown, neither barriers nor TCC layers are useful for such
development in \rel, that is, when communication is reliable but not
FIFO\@. In another paper~\cite{EM2005b}, we use essentially the same
framework to investigate safe composition in models with message
duplication or loss. Barriers and TCC layers are also absent in those
models. The framework introduced here is used to define two more
notions, namely fitting after and separating, that are more readily
applicable in those models.\footnote{We say that \emph{$P$ fits after
    $Q$} if $\chop Q P \refines_\Gamma \chop Q \chop P$. Program
  \emph{$S$ separates $P$ from $Q$} if $\chop P\lay S\lay Q \chop
  \refines_\Gamma P \chop S \chop Q$.}
We illustrate our approach by applying it to the case of \rel.
Notably, the approach allows for seamless composition of programs
without need for translation or headers.

The central notion introduced and explored in this paper is that of
one program \emph{sealing} another. Larger programs can be composed
from smaller ones provided each smaller program seals its predecessor.
For instance, recall that $\valt{}\lay\valt[j\to i]{}$ seals itself in
\rel. Lemma~\ref{l:seal.algebra}.\ref{it:l:seal.algebra.rpt} can be
used to show that a program of the form
\begin{math}
  \kw{while}\;\True\;\kw{do}\;\valt{}\lay\valt[j\to i]{}\;\kw{od}
\end{math}
can serve to transmit a sequence of values from $i$ to $j$ in \rel.
Indeed, if the return messages from $j$ to $i$ are not merely
acknowledgments, it can perform sequence exchange.
The notion of sealing in \rel is shown to be intimately related to
Lamport causality. Based on this connection, we devise efficient
algorithms for deciding and constructing seals for the class of
straight-line programs.

\subsection*{Acknowledgment}
\label{sec:acknoledgment}

We would like to thank Manuel Chakravarty, Yael Moses, and Ron van der
Meyden for helpful comments on preliminary versions of this paper.
Special thanks to Elena Blank for an observation that led to a
simplification of the notation.

\bibliographystyle{alpha}
\bibliography{strings,al,k,skript,kai,ron,concurrency,collections,security}
\label{p.last}
\appendix

\section{Algorithms}
\label{app:algo}
\begin{codebox}
  \Procname{$\proc{Program-Graph}(P)$}
  \li $V, E \gets \bigcup_{i\in\agents}\{\Fd_i,\Ld_i\}, \emptyset$
  \>\>\>\>\>\>\>\>\>\>\Comment First and last dummy nodes for each process
  \li $f \gets \lambda i:\agents.\Fd_i$
  \>\>\>\>\>\>\>\>\>\>\Comment Book keeping for local precedence
  \li \Comment Add sends and receives with local precedence
  \zi \For $e$ in $P$ from left to right where $e$ is of the form $\snd[i\to j]{}$ or $\rcv[i\from j]{}$
  \zi     \Do $V, E, f(i) \gets V\cup\{e\}, E\cup\{(f(i),e)\}, e$
  \End
  \li \Comment Add precedence between last $i$-event and $i$'s last
  dummy node
  \zi \For $i\in\agents$
  \zi     \Do $E\gets E\cup\{(f(i), \Ld_i)\}$
  \End
  \li \Comment Add precedence between FIFO matching sends and receives
  \zi \For $e\in V$ the $k$'th event in $P$ of the form $\snd[i\to j]{}$ for some $i,j,k$
  \zi     \Do $E \gets E\cup\{(e,e')\}$ where $e'$ is the $k$'th
  $\rcv[i\from j]{}$ event in $P$
  \End
  \li \Return $(V, E)$
\end{codebox}
\begin{codebox}
  \Procname{$\proc{Deadlock-Free}(P)$}
  \li $V, E \gets \proc{Program-Graph}(P)$
  \li \Return $\exists$ cycle in $E$
\end{codebox}
\begin{codebox}
  \Procname{$\sig(P)$}
  \li $V, E \gets \proc{Program-Graph}(P)$
  \li $E \gets E^+$\>\>\>\>\>\>\>\>\>\>\Comment Add irreflexive transitive closure
  \li \Comment Remove all but minimal sends and maximal receives on
  open channels
  \zi $V \gets V\setminus\setcpr{e}{\text{$e$ is a $\snd{}$ event
      preceded by another such send or $\Fd_j$}}$
  \zi $V \gets V\setminus\setcpr{e}{\text{$e$ is a $\rcv{}$ event that
      precedes another such receive or $\Ld_i$}}$
  \li \Return $(V, E\cap V^2)$
\end{codebox}
\newcommand{\safechan}{\id{safe}}
\begin{codebox}
  \Procname{$\proc{Is-Seal}(P,Q)$}
  \li $V_P, E_Q \gets \sig(P)$
  \li $V_Q, E_Q \gets \sig(Q)$
  \li \For $(i,j) \in \agents^2\setminus\Id_\agents$ s.t.~$\rcv[j\from i]{} \in V_P$
  \li     \Do $e\gets \begin{cases}
                        \snd{} & \text{if }\snd{} \in V_Q\\
                        \Ld_i & \text{otherwise}
                      \end{cases}$
  \li         $\safechan \gets \False$
  \li         \For $k \in \agents \setminus \{i\}$
  \li             \Do $\safechan \gets \safechan \Or
                      ((\rcv[j \from i]{}, \Ld_k) \in E_P \And
                       (\Fd_k, \snd[i\to j]{}) \in E_Q)$
                  \End 
  \li         \If $\Not\safechan$
  \li             \Then \Return $\False$
                  \End 
              \End 
  \li \Return $\True$
\end{codebox}
\begin{codebox}
  \Procname{$\proc{Closed-Channels}(P)$}
  \li $V, E \gets \agents, \agents^2\setminus\Id_\agents$
  \li $V', E' \gets \sig(P)$
  \li \For $(i,j) \in E$
  \li     \Do \If $\rcv[j\from i]{} \in V'$ and $(\rcv[j\from i]{}, \Ld_i) \notin E'$
  \li             \Then $E \gets E \setminus \{(i,j)\}$
                  \End
          \End
  \li \Return $(V, E)$
\end{codebox}
\begin{codebox}
  \Procname{$\proc{Signature-Compose}(V_P,E_P,V_Q,E_Q)$}
  \li \Comment Sequentially compose the two signatures
  \zi $V \gets \lbsetcpr{e^{(X)}}{e\in V_X \And X\in\{P,Q\}}$
  \zi $E \gets \lbsetcpr{(e^{(X)},f^{(Y)})\in V^2}{X=Y\And (e,f)\in
    E_X} \cup \lbsetcpr{(\Ld_i^{(P)}, \Fd_i^{(Q)})}{i\in\agents}$
  \li $E \gets E^+$
  \li \Comment Remove dummy nodes between the two signatures
  \zi $V \gets V \setminus \lbsetcpr{\Ld_i^{(P)}}{i\in\agents}
  \setminus \lbsetcpr{\Fd_i^{(Q)}}{i\in\agents}$
  \li \Comment Remove all but the first sends and last receives
  \zi $V \gets V \setminus \setcpr{e^{(Q)}\in V}{e^{(P)} \in V \And e=\snd{}}
  \setminus \setcpr{e^{(P)}\in V}{e^{(Q)} \in V \And e=\rcv{}}$
  \li \Comment Remove sends and receives on closed channels
  \zi $V \gets V \setminus \setcpr{e^{(Q)}\in V}{(\Fd_j,e^{(Q)}) \in E \And e=\snd{}}
  \setminus \setcpr{e^{(P)}\in V}{(e^{(P)}, \Ld_j) \in E \And e=\rcv{}}$
  \li rename by dropping superscripts $(X)$
  \li \Return $(V,E\cap V^2)$
\end{codebox}

\end{document}